\documentclass[intlimits,twoside,a4paper]{article}

\usepackage{amsmath,amssymb}
\usepackage{graphicx,psfrag}

\usepackage[T2A]{fontenc} %T2A
\usepackage[cp1251]{inputenc} %cp1251

\usepackage[eqsecnum]{cmpj2}
\issue{2014}{17}{2}{23602}
\doinumber{10.5488/CMP.17.23602}

\RequirePackage{color}

\title[Scaling functions and amplitude ratios for the Potts model %on an uncorrelated scale-free networks
]%
{Scaling functions and amplitude ratios for the Potts model on an
uncorrelated scale-free network}
\author[M. Krasnytska]{M. Krasnytska\refaddr{label1,label2}}
\addresses{
\addr{label1} Institute for Condensed Matter Physics of the National Academy
of Sciences of Ukraine, \\
1~Svientsitskii~St., 79011 Lviv, Ukraine
\addr{label2}
Institut Jean Lamour, CNRS/UMR 7198, Groupe de Physique Statistique,
 Universite de Lorraine, \\
 BP 70239, F-54506 Vand\oe uvre-l\'es-Nancy Cedex,
 France
}

\date{Received February 4, 2014, in final form February 19, 2014}
\authorcopyright{M. Krasnytska, 2014}

\begin{document}

\maketitle

\begin{abstract}
We study the critical behaviour of the $q$-state Potts model on an
uncorrelated scale-free network having a power-law node degree
distribution with a decay exponent $\lambda$. Previous data show
that the phase diagram of the model in the $q,\lambda$ plane in the
second order phase transition regime contains three regions, each
being characterized by a different set of critical exponents. In
this paper we complete these results by finding analytic expressions
for the scaling functions and critical amplitude ratios in the above
mentioned regions. Similar to the previously found critical
exponents, the scaling functions and amplitude ratios appear to be
$\lambda$-dependent. In this way, we give a comprehensive
description of the critical behaviour in a new  universality
class.
\pacs 64.60.aq, 64.60.fd, 64.70.qd, 64.60.ah
\end{abstract}

\section{Introduction}

The concept of universality plays a fundamental role in the theory
of critical phenomena \cite{Fisher67, Kadanoff67, Domb}.
 A lot of systems manifest similar behaviour near
the critical point. The universality class does not depend on the
local parameters but on the global ones, i.e., dimensionality, symmetry,
nature of interaction, etc. If several systems are in the same
universality class, they share, besides the values of the critical
exponents, identical critical amplitude ratios and scaling functions
\cite{Privman91}.

The goal of our study is to analyze an universal content of the
critical behaviour of the Potts model on a scale-free network near
the critical point, in particular, to quantify it in terms of
scaling functions and universal amplitude ratios. A lot of studies
were devoted to the analysis of the critical behaviour of spin
models on complex networks \cite{Dorogovtsev08}. In this case, the
disorder of an underlying structure is modelled in terms of a random
graph. In the present work we will consider the $q$-state Potts
model on uncorrelated scale-free network having a power-law node
degree distribution with exponent $\lambda$. Similar to the lattice
systems, for the Potts model on the uncorrelated scale-free networks
one may observe either the 1st or the 2nd order phase transition.
However, now the order of the phase transition depends, besides the
$q$ value, on the node degree distribution decay exponent $\lambda$
\cite{Krasnytska13,Dorogovtsev04,Igloi02}. The second order phase
transition regime is characterized by power law dependencies of
thermodynamic functions as functions of temperature and magnetic
field in the vicinity of critical point. Critical exponents governing
this transition depend on $\lambda$, which plays the role of a
global variable for models on a network, like the dimension $d$ in the
case of a lattice.

Depending on the particular value of $q$, the Potts model has been suited to describe various
real and model systems. Besides the Ising model at $q=2$, it also describes
 percolation at $q\rightarrow 1$ \cite{Fortiun69,Giri77}. The spanning
treelike percolation with a geometric phase transition is described by a zero-state
$q=0$ Potts model \cite{Stephen76}. Also in $q=0$ limit,
the Potts model can be used for a description of the Kirchhoff`s rules via the resistor network models
\cite{Fortiun72}. Subsequently, there has been shown the equivalence between the zero-state
Potts model and Abelian sandpile models in case of arbitrary
finite graphs \cite{Majumdar92}. Sandpile models describe processes
in neural networks, fracture, hydrogen bonding in liquid water.
Another particular case of Potts model at $q=1/2$ is a spin glass model
\cite{Aharony78,Aharony79}. The case of $0\leqslant q<1$ is used to describe
gelation and vulcanization processes in branched polymers
\cite{Lubensky78}. Other examples concern the application of the Potts model for
larger values of $q$. Three-component $q=3$ Potts model is used to
describe a cubic ferromagnet with three axes in a diagonal magnetic
field \cite{Mukamel76}, an adsorption of 4He atoms on graphite in
two dimensions \cite{Alexander75}, transition of helium films on
graphite substrate \cite{Bretz77}, etc. The $4$-state Potts model
also describes the effect of absorbtion on surfaces \cite{Domany77}.
The Potts model at large $q$ is used to simulate the processes of
intercellular adhesion and cancer invasion \cite{Turner02},
see also \cite{Laanait91}.

In this paper we will complete the analysis of the Potts model on an
uncorrelated scale-free network \cite{Krasnytska13} by calculating
scaling functions and universal amplitude ratios. Recently,
\cite{vonFerber11} the scaling functions and universal amplitude
ratios were obtained for the Ising model on a scale free network.
Here we will generalize these expressions for the Potts model case.

The structure of this paper is as follows. In the next section we
write down the main relations of the scaling theory,  expressions for
thermodynamic functions in a scaling form and universal amplitude
ratios. Section \ref{III} is a short overview of our previous work,
where the critical behavior of the Potts model on uncorrelated
scale-free network was considered. In particular, it was shown that
the phase diagram in the second order phase transition regime
contains three regions, each being characterized by a different set
of critical exponents. In section \ref{IV}, we complete these
results by finding analytic expressions for the scaling functions
and critical amplitude ratios in the above mentioned regions.
Similar to the previously found critical exponents, the scaling
functions and amplitude ratios appear to be $\lambda$-dependent. In
this way, we give a comprehensive description of the critical
behaviour in a new universality class. Analytic expressions are
summarized in table~\ref{tab2}. In the last section we summarize the obtained
results.

\section{Main relations of scaling theory}\label{II}

In this paper we will be interested in the universal features of a
system that are manifested in the vicinity of the critical (i.e., second
order phase transition) point. Critical exponents, that govern
the power-law behaviour of different observables near the critical point
belong to such characteristics. Temperature
driven phase transition into magnetically ordered state being taken for definiteness, one
observes the power-law asymptotics near the critical point $T=T_\mathrm{c},\,
h=0$ \cite{Stanley99,Privman91}. In particular, at $h=0$ the
(dimensionless) order parameter $m$, isothermal susceptibility
$\chi_T$, specific heat $c_h$ and magnetocaloric coefficient $m_T$
are\footnote{The magnetocaloric coefficient is defined by the mixed
derivative of the free energy over magnetic field and temperature,
$m_T=-T(\partial m/\partial T)_h$. It measures the heat released by
the system upon an isothermal increase of the magnetic field due to
the magnetocaloric effect (see, e.g., \cite{vonFerber11} and
references therein). } power law functions of
$\tau={|T-T_\mathrm{c}|}/{T_\mathrm{c}}$:
\begin{equation}\label{15}
m=B_-\tau^\beta,\qquad \chi_T =\Gamma_\pm \tau^{-\gamma}, \qquad
c_h=\frac{A_\pm}{\alpha}\tau^{-\alpha},\qquad m_T=B_T^{\pm} \tau^{-\omega}
\qquad \text{at} \qquad h=0.
\end{equation}
Here, indices $\pm$ refer to the way the critical temperature is
approached, $T-T_\mathrm{c}\to 0^{\pm}$. In turn, directly at $T=T_\mathrm{c}$ (i.e.,
$\tau=0$) the following power law field dependencies hold:
\begin{equation}\label{16}
m=D_\mathrm{c}^{{-1}/{\delta}}h^{1/\delta}, \qquad \chi=\Gamma_\mathrm{c}
h^{-\gamma_\mathrm{c}}, \qquad c_h=\frac{A_\mathrm{c}}{\alpha_\mathrm{c}}h^{-\alpha_\mathrm{c}},
\qquad m_T=B^c_T h^{-\omega_\mathrm{c}} \qquad \text{at}
\qquad \tau=0.
\end{equation}
The above formulas (\ref{15}), (\ref{16}) introduce critical
exponents and critical amplitudes that we are interested in in this
study. Unlike the critical exponents, the critical amplitudes are
non-universal, being dependent on the microscopic features of the system.
However, their certain combinations appear to be universal as well
\cite{Privman91}. In particular, in this study we will be interested
in the following universal critical amplitude  ratios:
\begin{equation}\label{12}
R_\chi^\pm=\Gamma_\pm D_\mathrm{c} B_-^{\delta-1}\,, \quad R_\mathrm{c}^\pm=\frac{A_\pm \Gamma_\pm}{\alpha B_-^2}\,,\quad
R_A=\frac{A_\mathrm{c}}{\alpha_\mathrm{c}} D_\mathrm{c}^{-(1+\alpha_\mathrm{c})}B_-^{-2/\beta}\,,
\quad A_+/A_-\,, \quad \Gamma_+/\Gamma_-\,.
\end{equation}

The above quoted power law scaling in the behaviour of various thermodynamic functions,
 universality and scaling relations between critical exponents and
amplitude ratios are the manifestations of special properties of the
thermodynamic  potential in the vicinity of critical point. In
particular, the scaling hypothesis for the Helmholtz free energy
$F(\tau,m)$ states that this thermodynamic potential is a generalized
homogeneous function \cite{Stanley72} and can be written as follows:
\begin{equation}\label{0}
F(\tau,m)\approx \tau^{2-\alpha} f_{\pm}(x),
\end{equation}
with the scaling variable $x=m/\tau^\beta$ and scaling function $f_\pm(x)$, signs
$+$ and $-$ correspond to $T>T_\mathrm{c}$ and $T<T_\mathrm{c}$, respectively. The principal content of equation (\ref{0})
is that $F(\tau,m)$ as a function of two variables can be mapped onto a single variable scaling function $f_\pm(x)$.
It may be shown that all thermodynamic potentials are generalized homogeneous functions, provided
one of them possesses such a property \cite{Stanley72}.

Based on the expression for the free energy one can also represent
the thermodynamic functions in terms of appropriate scaling functions.
In particular, magnetic and entropic equations of state read:
\begin{eqnarray} \label{97}
 h(m,\tau) &=& \tau^{\beta \delta}H_\pm (x), \\ \label{99}
S(m,\tau) &=& \tau^{1-\alpha} {\cal S}(x),
\end{eqnarray}
with the scaling functions $H_\pm (x)$ and ${\cal S}(x)$. In turn,
the scaling functions for the heat capacity, isothermal
susceptibility, and magnetocaloric coefficient are defined via (see
e.g.,~\cite{vonFerber11}):
\begin{eqnarray}  \label{100}
 c_h(m,\tau) &=&(1\pm \tau)\tau^{-\alpha} {\cal C_\pm}(x),
  \\ \label{101}
 \chi_T(m,\tau) &=&\tau^{-\gamma}\chi_\pm(x),  \\ \label{102}
 m_T(m,\tau) &=& (1\pm\tau)\tau^{\beta-\gamma}{\cal M}_\pm(x).
\end{eqnarray}

Scaling functions are reachable in experiments and MC simulations. Together with critical exponents and
critical amplitude ratios they constitute quantitative characteristics of a given universality class.
In the rest of this paper we will complete the previous description of the critical behaviour of the
Potts model on an uncorrelated scale-free network by calculating its amplitude ratios and scaling functions
in the vicinity of the second order phase transition.

\section{Potts model on an uncorrelated scale-free network} \label{III}

 The $q$-state Potts model can be considered as one of the possible
 generalizations of the Ising model, where the spin variable
 can have $q$ possible states \cite{Wu82}. The Potts model Hamiltonian reads:
  \begin{equation}\label{2}
-H=\frac{1}{2}\sum_{i,j}J_{ij}\delta _{n_i,n_j}+h\sum _i
\delta_{n_i,0}\, , \qquad (n_i=0,1,\ldots, q-1),
  \end{equation}
here $q$ is the number of Potts states, $h$ is a local external
magnetic field directed along the $0$-th component of the Potts
variable $n_i$ (the Potts state on the node $i$). We consider the case
where all spins are located on the nodes of a random graph (complex network)
and are connected with each other in an appropriate way. The latter is determined by the adjacency
matrix  $J_{ij}$ with the elements $J_{ij}=1$ if there exists a link between the nodes $i$ and $j$
and $J_{ij}=0$ otherwise. One of the important characteristics of a network is its node degree
distribution $P(k)$: a probability that the randomly chosen
node has a degree (number of links) $k$. We will consider the case of
Potts model on an uncorrelated
scale-free network with a power-law node degree distribution:
\begin{equation}
 P(k)=c_\lambda k^{-\lambda},
\end{equation}
here, $c_\lambda$ is a normalization constant and $\lambda$ is the
exponent of decay. The absence of correlations within the given link
distribution means that the probability to create a link between two
nodes is linearly proportional to their node degrees. Furthermore,
one may consider the case of an annealed network, when the network
configuration is fluctuating under the constraint of a given node degree
distribution, see, e.g., \cite{annealed,annealed2}. Alternatively, the links
between the nodes may be randomly distributed but remain fixed in a given
configuration, the so-called configurational model, see
\cite{Dorogovtsev08,Dorogovtsev02}. The latter situation corresponds to the
quenched case and is usually more complicated for analytical
treatment.

The critical behaviour of the Potts model on an uncorrelated scale-free network has
been considered in references \cite{Krasnytska13,Dorogovtsev04,Igloi02}. It was found that the
phase diagram of the model is uniquely defined by two parameters: the number of  Potts
states $q$ and the node degree distribution exponent $\lambda$. Here, we will complete
the calculations of \cite{Krasnytska13} where a comprehensive list of critical exponents
governing the behaviour of thermodynamic functions in the second order phase transition
regime was found.  The results of \cite{Krasnytska13} are exact for an annealed network and
correspond to the mean field treatment of the quenched case. Our starting point will be the expression
for the Helmholtz free energy obtained in reference \cite{Krasnytska13} for different $q$. For non-integer $\lambda>3$,
the free energy reads:
\begin{equation}\label{aa}
F(\tau,M)=a_1\tau M^2+a_2 M^{\lambda-1} +  \sum_{i=3}^{[\lambda-1]}a_i M^{i} + O(M^{[\lambda]}) \, ,
\end{equation}
here, $M$ is magnetization, $a_i$ are non-universal coefficients,
their explicit form is given in \cite{Krasnytska13} and $[\lambda]$
is the integer part of $\lambda$. Note that the power law polynomial
form (\ref{aa}) holds for the Helmholtz potential for non-integer
$\lambda$ only. Logarithmic corrections appear in the case of
integer values of $\lambda$. As we discuss below, this will
lead to the changes in the critical behaviour at $\lambda=4$  and
$\lambda=5$.

\begin{figure}[!t]
\begin{center}
\includegraphics[width=0.475\textwidth]{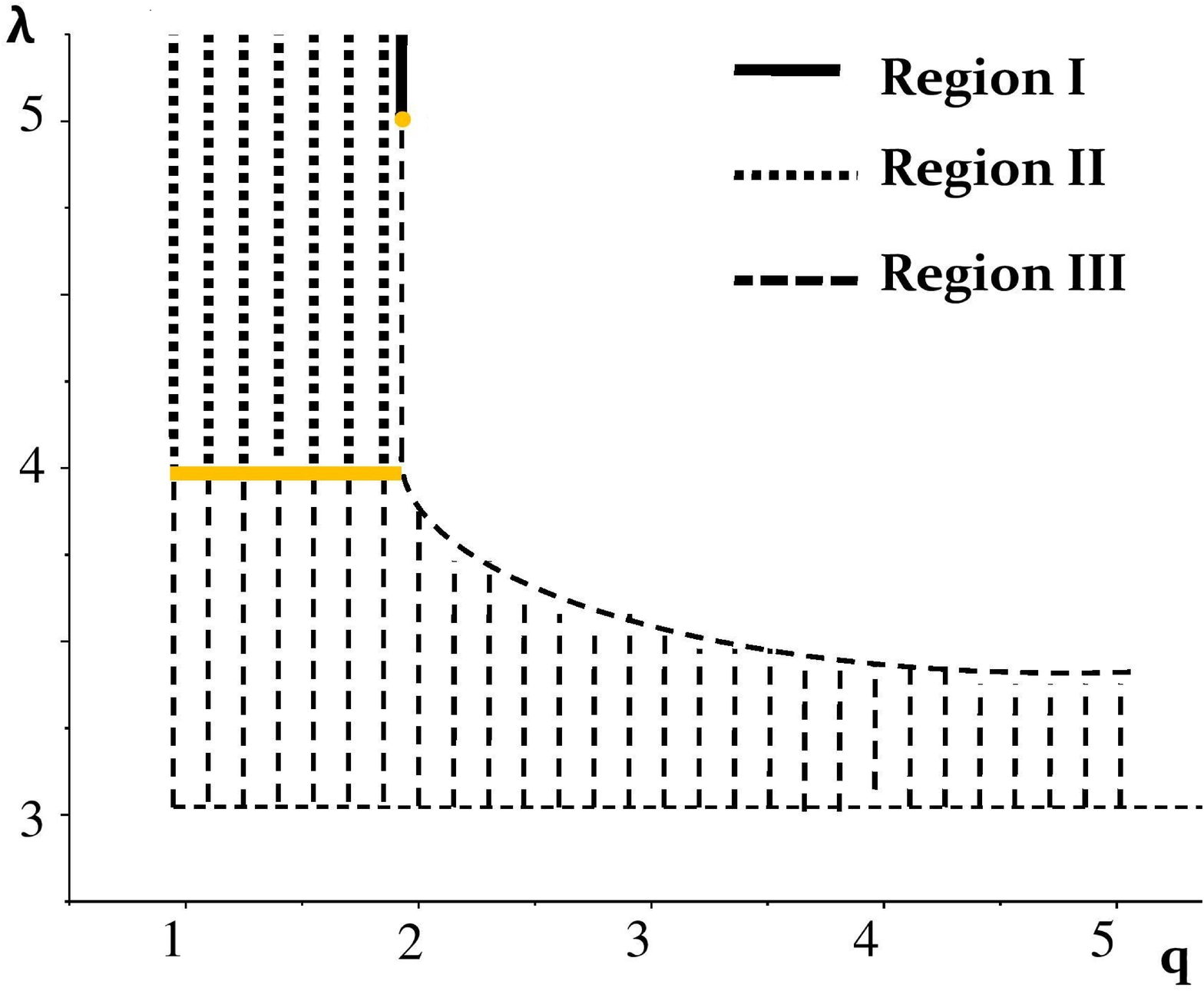}
\caption{\label{Fig1} (Color online) Phase diagram of the Potts model on an
uncorrelated scale-free network. The area of the second order phase
transition is shown. Three regions correspond to three different
universality classes. Logarithmic corrections to scaling appear for
the values of $q$ and $\lambda$ shown by a light line  and a light
disc (brown online). See the text for a more detailed discussion.}
\end{center}
\end{figure}

Figure~\ref{Fig1} generalizes the information about the critical
behavior of the Potts model on an uncorrelated scale-free network in
the form of a phase diagram in the $q-\lambda$ plane
\cite{Igloi02,Krasnytska13}. It has been  shown that for
$\lambda\leqslant 3$, the system remains in the ordered state for any
finite temperature \cite{Krasnytska13,Dorogovtsev04,Igloi02}. For
$\lambda > 3$, the phase transition may be of the first or second
order, depending on the  specific values of $q$ and $\lambda$. The
first order phase transition occurs at the values of $q$ and
$\lambda$ that belong to the blank region above the dashed curve,
$q>2$, $\lambda>\lambda_\mathrm{c}(q)$. Of the main interest for us will be the
second order phase transition regime. This corresponds to three
different regions shown in figure~\ref{Fig1} that belong to three
different universality classes. Region I, $\lambda>5$, $q=2$ (black
solid line in figure~\ref{Fig1}) is governed by the Ising mean field
critical exponents. Region II, $\lambda>4$, $1\leqslant q<2$ (dotted area
in the figure) is governed by the percolation mean field critical
exponents. Region III ($3<\lambda<5$, $q=2$; $3< \lambda<4$, $1\leqslant
q<2$; $3<\lambda\leqslant \lambda_\mathrm{c}(q)$, $q>2$) (dashed area in the
figure) is characterized by the non-trivial $\lambda$-dependency of
the critical exponents. Values of the critical exponents in all
three regions are collected in table~\ref{tab1}.

\begin{table}[!t]
\caption{Critical indices of the Potts model on an uncorrelated scale-free network in three
different regions of $q$ and $\lambda$ values, see figure~\ref{Fig1}.
One recovers the results for the Ising model ($q=2$ \cite{Leone02,Dorogovtsev03}) and for percolation ($q=1$) \cite{Cohen02}. \label{tab1}  }
\vspace{2ex}
\begin{center}
\begin{tabular}{|l|c|c|c|c|c|c|c|c|}
\hline  &$\alpha$ & $\alpha_\mathrm{c}$ & $\beta$ &  $\delta$ & $\gamma$ &
$\gamma_\mathrm{c}$ &
$\omega$ & $\omega_\mathrm{c} $ \\
\hline
region I & 0    & 0 &  1/2 &  3 &   1 &  2/3 &  1/2 &  1/3  \\
region II & $-1$    & $-1/2$ &  1 &  2 &   1 &  1/2 &  0 &  0  \\
region III & $\frac{\lambda-5}{\lambda-3} $ &
$\frac{\lambda-5}{\lambda-2}$ & $ \frac{1}{\lambda-3}$ & $\lambda-2$
&  $1$ &  $ \frac{\lambda-3}{\lambda-2}$ &
$\frac{\lambda-4}{\lambda-3}$ &
 $\frac{\lambda-4}{\lambda-2}$  \\[1ex]
\hline
\end{tabular}
\end{center}
\end{table}

As it was mentioned above, logarithmic terms appear in the free
energy at the integer values of $\lambda$. In turn, this leads to
the appearance of logarithmic corrections \cite{Kenna06,Kenna06',
Berche13} to the power-law scaling dependencies (\ref{15}),
(\ref{16}) of thermodynamic functions at $\lambda=5$ for the
Ising model ($q=2$) and at $\lambda=4$ for $1\leqslant q<2$
\cite{Krasnytska13}. Values of $q$ and $\lambda$ where  the
thermodynamic functions are governed by power-law singularities
enhanced by the logarithmic corrections are shown in figure~\ref{Fig1}
by the light solid line and light disc (brown online).

In the forthcoming section we will be interested in the critical
behaviour in the regions  of the second order phase transition with
the power law scaling. In particular, we will complete a quantitative
description of three universality classes found in regions I, II and
III (see figure~\ref{Fig1}) by calculating, in addition to the
critical exponent, the scaling functions and amplitude ratios.

\section{Critical amplitude ratios and scaling functions} \label{IV}

The expression of the free energy of the Potts model on an uncorrelated
scale-free network (\ref{aa}) will be a starting point for the
analysis of the critical amplitude ratios and scaling functions.
Passing to the dimensionless energy $f(m, \tau)$ and dimensionless
magnetization $m$ and leaving the leading order contributions for small
values of $m$, we can present (\ref{aa}) in three different regions
of the phase diagram (figure~\ref{Fig1}) in the following form:
\begin{align}
 \label{880}
f(m,\tau) &=  \pm \frac{\tau}{2}m^2 + \frac{1}{4} m^4, & &\text{(Region I)} ,\\
%\end{equation}
%\begin{equation}
\label{881}
 f(m,\tau) &=  \pm \frac{\tau}{2}m^2 +
\frac{1}{4} m^3, & &\text{(Region II)} , \\
%\end{equation}
%\begin{equation}
\label{882}
f(m,\tau) &= \pm \frac{\tau}{2} m^2 + \frac{1}{4}m^{\lambda-1},
& &\text{(Region III)} ,
 \end{align}
the signs $\pm$ here and in what follows refer to the temperatures
above and below the critical point $T_\mathrm{c}$. Note that the positive
sign of the second terms in (\ref{880})--(\ref{882}) is due to the
fact that coefficients $a_2$, $a_3$, and $a_4$ in (\ref{aa})  are
positive definite in the regions III, II, and I,
correspondingly. With the expressions for the free energy at hand it
is straightforward to write down the equation of state and to derive
the thermodynamic functions. The magnetic and entropic equations of
state in the dimensionless variables $m$ and $\tau$ read:
\begin{equation}\label{eqs}
h(m,\tau)=\left.\frac{\partial f(m,\tau)}{\partial m} \right|_\tau \, , \qquad s(m,\tau)=\left.\mp \frac{\partial f(m,\tau)}{\partial \tau}
\right|_m \, .
\end{equation}

Written explicitly in different regions of $q$, $\lambda$ the magnetic equation of state attains the following form:
\begin{align}
\label{8800}
h &= m^3\pm \tau m, & & \text{(Region I)} , \\
\label{8810}
h &= \frac{3}{4} m^2\pm \tau m, & &\text{(Region II)} , \\
\label{8820}
h &= \frac{\lambda-1}{4} m^{\lambda-2}\pm \tau m, & &
\text{(Region III)}.
\end{align}
The entropic equation of state is obtained by a temperature
derivative at a constant magnetization $m$ while the explicit
$\tau$-dependency is the same in all expressions
(\ref{880})--(\ref{882}). Therefore, the equation keeps the same form
in all regions on $q$--$\lambda$ plane:
\begin{equation}\label{eq8820}
s=-m^2/2, \qquad\qquad \text{(Regions I--III)} .
\end{equation}

Thermodynamic functions $\chi_T$, $c_h$, and $m_T$ that characterize
the response on an external action are directly obtained from the above
equations of state. We do not present the explicit expressions here,
being rather interested in the corresponding critical  amplitude
ratios. The latter are given in table~\ref{tab2}. In the particular
case $q=2$, by these expressions we recover the formerly obtained
critical amplitude ratios for the Ising model on an uncorrelated
scale-free network \cite{Palchykov10,vonFerber11}, correcting at
$3<\lambda<5$ the expression for $R_A$ given in \cite{vonFerber11}\footnote{In paper \cite{vonFerber11}, using equation (\ref{12}) to find
$R_A$ the $\alpha$ exponent was substituted instead of $\alpha_\mathrm{c}$
into the power of $D_\mathrm{c}$.}. Together with the previously derived set of
critical exponents (see table~\ref{tab1}), our results for the
critical amplitude ratios quantify the universal features of
critical behaviour of the Potts model in the second order phase
transition regime.

\begin{table}[!t]
\caption{Scaling functions and critical
amplitude ratios for the Potts model on an uncorrelated scale-free network. \label{tab2} }
\vspace{1ex}
\begin{center}
\begin{tabular}{|c|c|c|c|}
\hline
              & Region I  &  Region II  &  Region III \\ \hline
$f_\pm(x)$   & $\pm \frac{x^2}{2} + \frac{x^4}{4}$ &$\pm \frac{x^2}{2} + \frac{x^3}{4}$ &$\pm \frac{x^2}{2} + \frac{x^{\lambda-1}}{4}$\\
$H_\pm(x)$   &  $x^3\, \pm \,x$ &$\frac{3}{4}x^2\, \pm \,x$&$\frac{\lambda-1}{4} x^{\lambda-2}\, \pm \, x$\\
${\cal S}(x)$   & $-x^2/2$ & $-x^2/2$ &$-x^2/2$\\
${\cal C_\pm}(x)$   &  $\frac{x^2}{3x^2 \pm 1}$&$\frac{ x^2}{3x/2\pm \, 1}$ &$\frac{ x^2}{(\lambda-1)(\lambda-2)x^{\lambda-3}/4\pm 1}$ \\
$\chi_\pm(x)$   &  $\frac{1}{3x^2\, \pm \, 1}$& $\frac{1}{3x/2\, \pm \, 1}$ &$\frac{1}{(\lambda-1)(\lambda-2)x^{\lambda-3}/4\, \pm \, 1}$ \\
${\cal M}_\pm(x)$  &  $\frac{x}{3x^2 \pm 1}$ &$\frac{ x}{3x/2\pm \, 1}$ &$\frac{ x}{(\lambda-1)(\lambda-2)x^{\lambda-3}/4\pm 1}$ \\
$A^+/A^-$   & $0$ & $0$ & $0$\\
$\Gamma^+/\Gamma^-$   & $2$ & $1$ & $\lambda-3$\\
$R^+_\chi$   & $1$ & $1$ & $1$\\
$R^-_\chi$   & $\frac{1}{2}$ & $1$ &$\frac{1}{\lambda-3}$\\
$R^+_\mathrm{c}$   & $0$ & $0$ &$0$\\
$R^-_\mathrm{c}$   & $\frac{1}{4}$ & $1$ &$\frac{1}{(\lambda-3)^2}$\\
$R_A$   & $\frac{1}{3}$ & $\frac{1}{2}$ &$\frac{1}{\lambda-2}$\\[1ex]
\hline
\end{tabular}
\end{center}
\end{table}

\begin{figure}[!t]
\begin{center}
\centerline{
\includegraphics[width=0.35\textwidth]{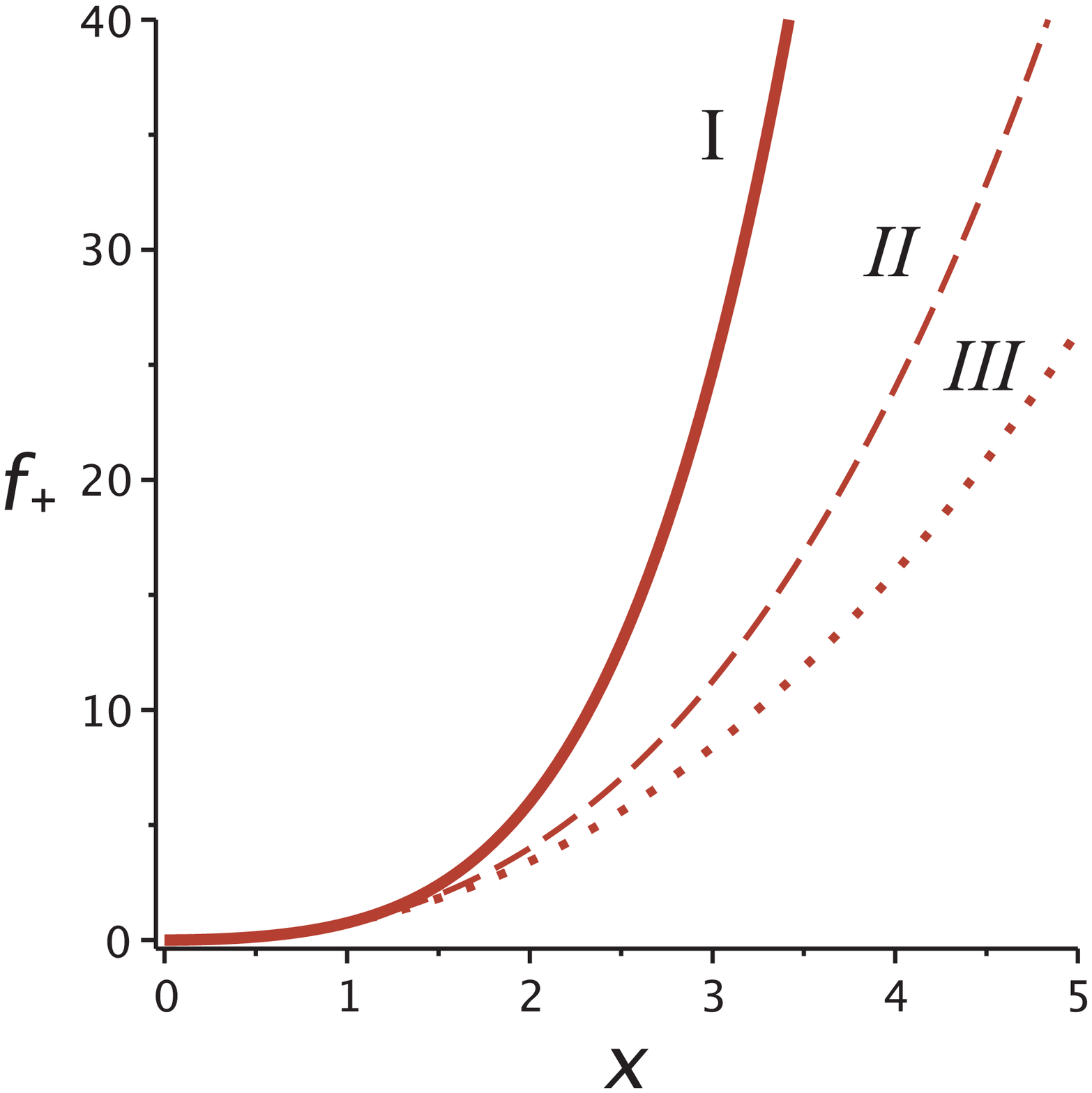}%\hspace{2em}
\hspace{1cm}
\includegraphics[width=0.35\textwidth]{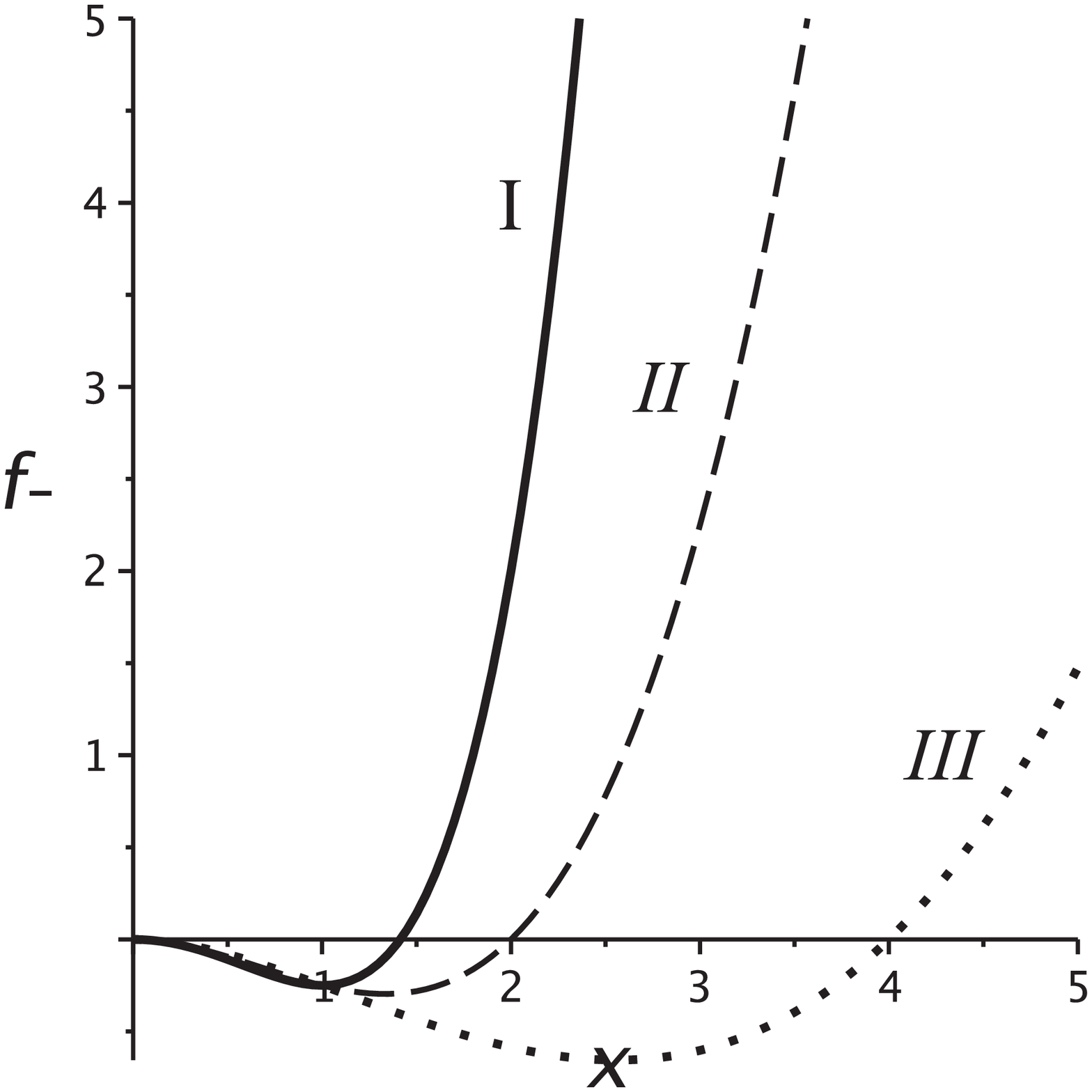}
}
\centerline{ \hspace{8mm}(a) \hspace{0.4\textwidth} (b)}
\caption{\label{Fig2}
(Color online) Limiting behaviour of the free energy scaling functions $f_+$ [(a), $T>T_\mathrm{c}$] and
$f_-$ [(b), $T<T_\mathrm{c}$]. The functions remain unchanged for $\lambda>5$, $q=2$ and
$\lambda>4$, $1\leqslant q < 2$ (solid and dashed curves I and II, correspondingly). For $q>2$ the phase transition
turns out to be of the first order at $\lambda>\lambda_\mathrm{c}(q)$. Dotted curves III: $q=4$,
$\lambda=\lambda_\mathrm{c}(4)\simeq 3.5$. See the text for more details.}
\end{center}
\end{figure}

Let us derive now the scaling functions for the free energy and other thermodynamic
functions. Using the definition (\ref{0}) and taking into account that the heat capacity
and the order parameter critical exponents $\alpha$, $\beta$ take on different values
in different regions of  the phase diagram figure~\ref{Fig1} (the formulas are given
in table \ref{tab1}) we can recast Helmholtz potential $F(\tau,m)$ in terms of the scaling
function $f_\pm(m/\tau^\beta)$. The explicit expressions for the scaling function in all three
regions of the phase diagram are given in table~\ref{tab2}. Typical behaviour of the free energy scaling
functions $f_+(x)$ and $f_-(x)$ is shown in figure~\ref{Fig2}~(a) and \ref{Fig2}~(b), correspondingly.

At any value of $q$, the scaling functions share a common feature: their curvature gradually increases
with an increase of $\lambda>3$. This happens up to some marginal value $\lambda=\lambda_\mathrm{c}$. The
marginal value $\lambda_\mathrm{c}$ is $q$-dependent. For $\lambda>\lambda_\mathrm{c}$ and  $1\leqslant q \leqslant 2$,
the scaling functions remain unchanged: their shape does not change with a further increase of $\lambda$.
The logarithmic corrections to scaling appear at $\lambda=\lambda_\mathrm{c}$ and the second order
phase transition holds in this case  for $\lambda>\lambda_\mathrm{c}$ as well \cite{Krasnytska13}, see figure~\ref{Fig1}.
Alternatively, for $\lambda>\lambda_\mathrm{c}$ and  $ q > 2$, the phase transition turns out to be of the first order.
Curves I of figure~\ref{Fig2} (plotted by solid lines) show the limiting behaviour of the scaling functions at $q=2$, $\lambda > 5$
[note that $\lambda_\mathrm{c}(q=2)=5$]: the functions remain unchanged for all $\lambda > 5$. Similar behaviour holds
for the case $1\leqslant q < 2$, the value of $\lambda_\mathrm{c}$, however, differs: $\lambda_\mathrm{c}(1\leqslant q < 2)=4$. This is
shown by curves II in the figure, plotted by dashed lines. Finally, curves III (dotted lines) for $q=4$  are one of
examples of the limiting behaviour of the scaling functions in the region $q>2$.

\begin{figure}[!b]
\begin{center}
\includegraphics[width=0.33\textwidth]{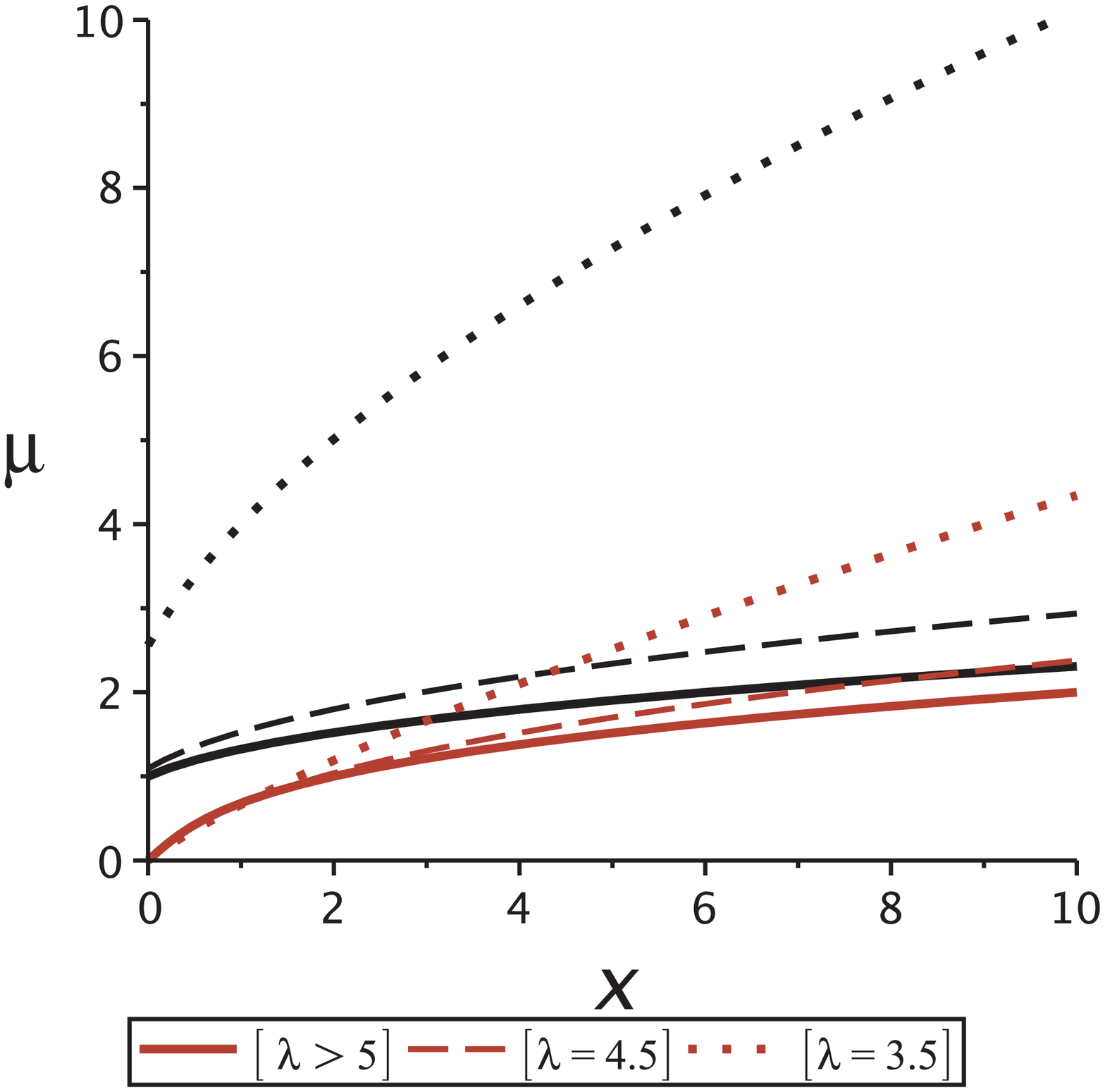} \hfill %\hspace{1em}
\includegraphics[width=0.33\textwidth]{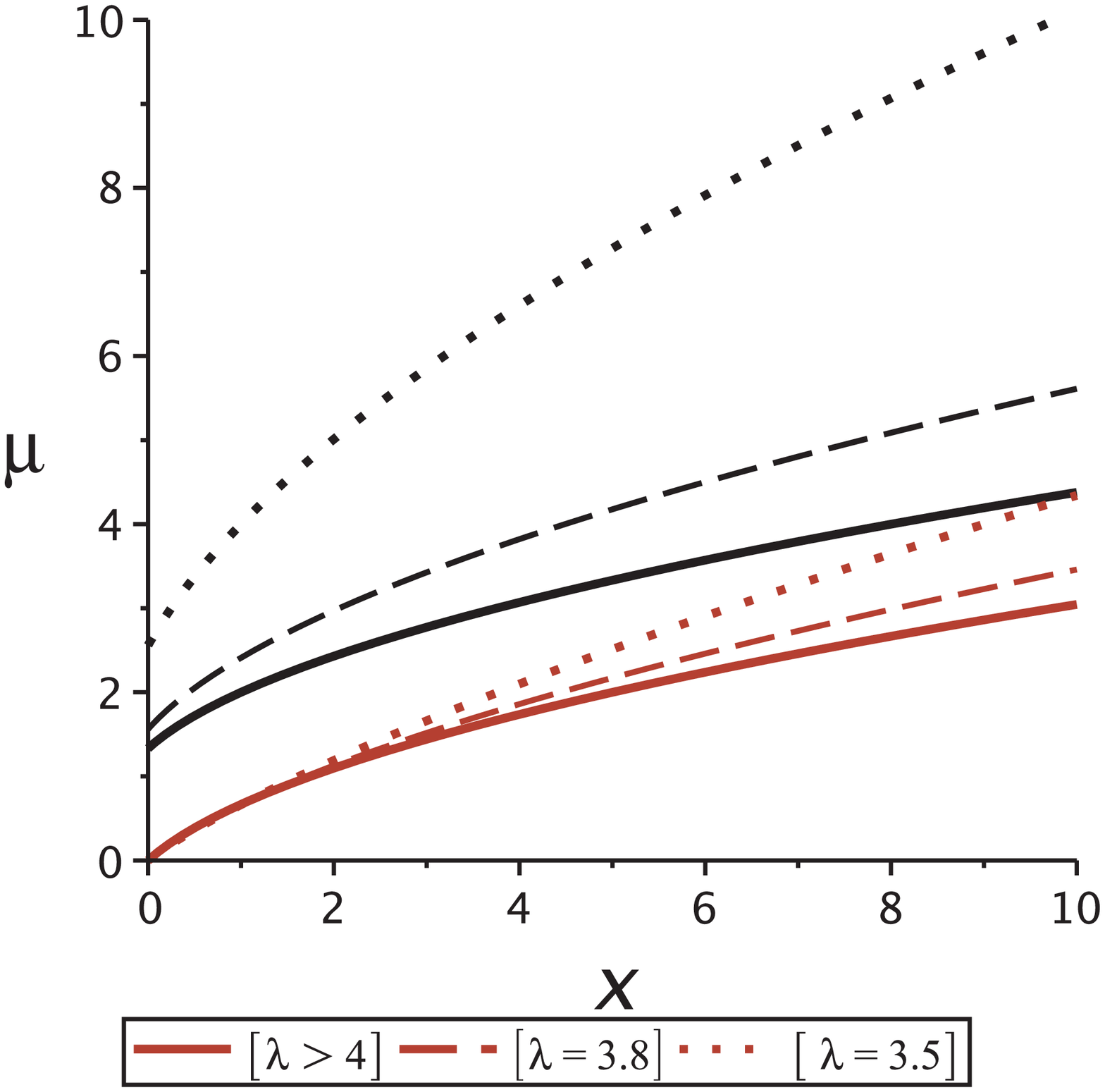} \hfill        %\hspace{1em}
\includegraphics[width=0.33\textwidth]{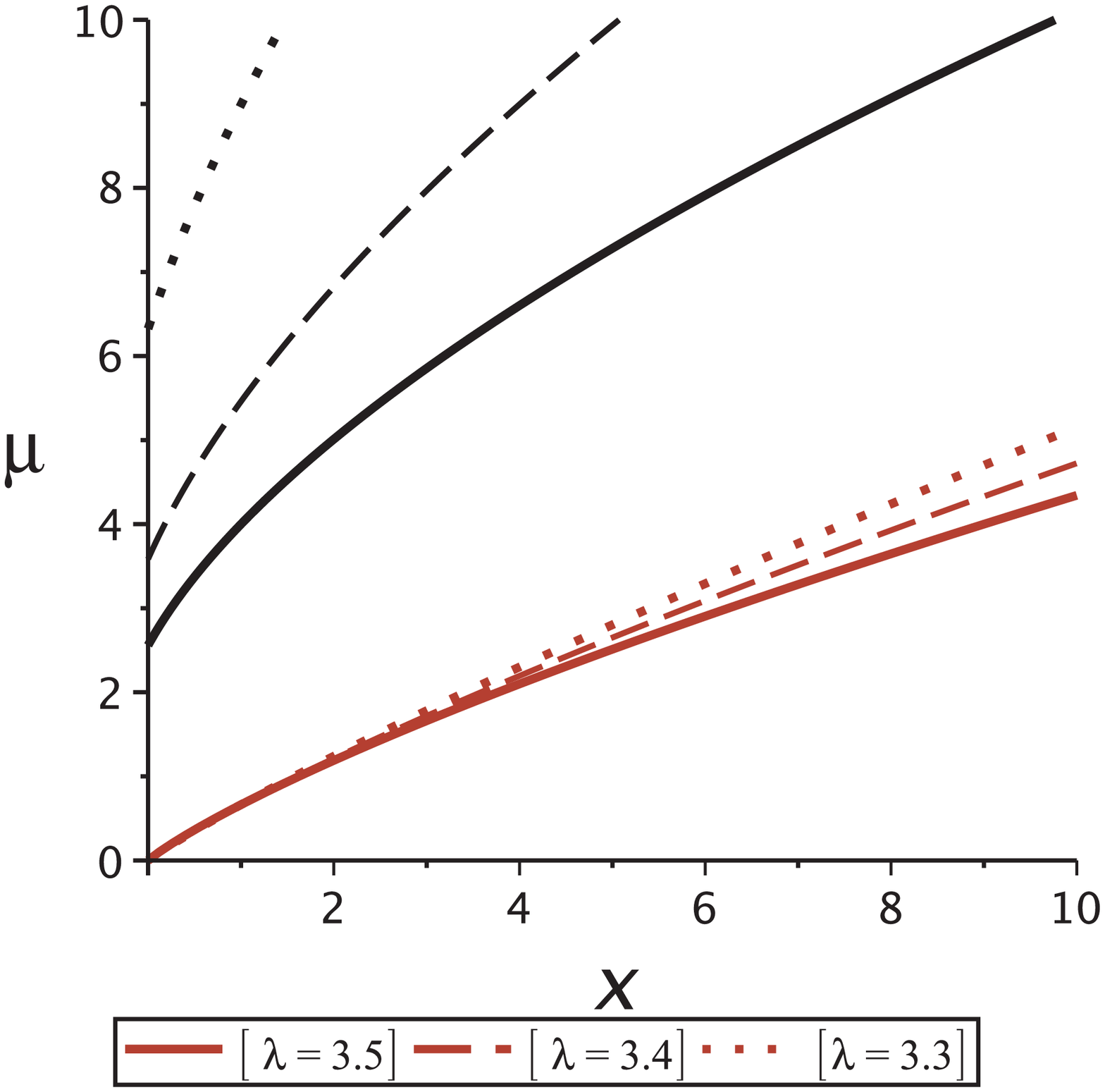}
\centerline{\hspace{5mm}{(a)} \hspace{15em} {(b)} \hspace{15em} {(c)}}
\caption{\label{Fig3} (Color online) Behaviour of the order parameter scaling
functions $\mu_+(x)$ (light curves, brown online) and $\mu_-(x)$
(black curves) for different values of $\lambda$ and $q$. {(a)}:
$q=2$, {(b)}: $1\leqslant q <2$, {(c)}: $q=4$. Values of $\lambda$
are shown in the figures.}
\end{center}
\end{figure}

Entropy scaling function ${\cal S} (x)$ is defined by (\ref{99}).
Using expression (\ref{eq8820}) for the entropy and taking into
account the values for the critical exponents $\alpha$ and $\beta$
given in table \ref{tab1} we arrive at the entropy scaling function
that remains unchanged in all regions on $q$--$\lambda$ plane:
${\cal S} (x)=-x^2/2$. There are different ways of representing the
magnetic equation of state. In the Widom-Griffiths representation
\cite{Griffiths67, Widom65} the magnetic equation of state can be
written in two equivalent forms:
\begin{equation}\label{wg}
h=m^\delta h_\pm \left(\tau/m^{1/\beta}\right) \, , \qquad
h=\tau^{\beta\delta} H_\pm \left(m/\tau^\beta\right) \, ,
\end{equation}
with scaling functions $h_\pm(x)$ and $H_\pm(x)$. Alternatively, in  Hankey-Stanley representation,
the magnetization is written as \cite{Stanley72}:
\begin{equation}\label{hs}
m=\tau^\beta \mu_{\pm}\left(h/\tau^{\beta\delta}\right) \,
\end{equation}
with the scaling function $\mu_\pm(x)$.

Starting from the magnetic equation of state given in regions I--III
by equations (\ref{8800})--(\ref{8820}), it is straightforward to arrive
at the scaling functions $H_\pm(x)$. We give the appropriate
expressions in table~\ref{tab2}. Subsequently, one can easily
rewrite these expressions to get appropriate $h_\pm$-functions.
Behaviour of the scaling functions $\mu_\pm(x)$ for different values
of $\lambda$ and $q$ is shown in figure~\ref{Fig3}. From  the explicit
form of the equation of state it is easy to evaluate the asymptotic
behaviour of the scaling functions. For $q=2$ and $\lambda>5$, one
gets $\mu_\pm(x)\sim x^{1/3}, \, x\to\infty$. The functions
tend to turn to infinity faster with a decrease
of $\lambda$:  $\mu_\pm(x)\sim x^{1/(\lambda -2)}, \, x\to\infty$
for $3<\lambda<5$. A similar feature is observed for the other
values of $q$. At $1\leqslant q<2$, $\lambda>4$ one gets $\mu_\pm(x)\sim
\sqrt{x}, \, x\to\infty$ and $\mu_\pm(x)\sim x^{1/(\lambda -2)}, \,
x\to\infty$ for $3<\lambda<4$. The last asymptotic behaviour
also holds for $q>2$ and $\lambda\leqslant \lambda_\mathrm{c}(q)$. Note that all
light curves (brown online) in figure~\ref{Fig3} start form the
origin: this corresponds to the absence of spontaneous magnetization
at $T>T_\mathrm{c}$. Correspondingly, the value of the scaling function
$\mu_-(x)$ at $x=0$ gives the spontaneous magnetization critical
amplitude $B_-$, equation (\ref{16}). As one can see in figure~\ref{Fig3}, the latter  increases with a decrease of $\lambda$.

\begin{figure}[!t]
\begin{center}
\includegraphics[width=0.33\textwidth]{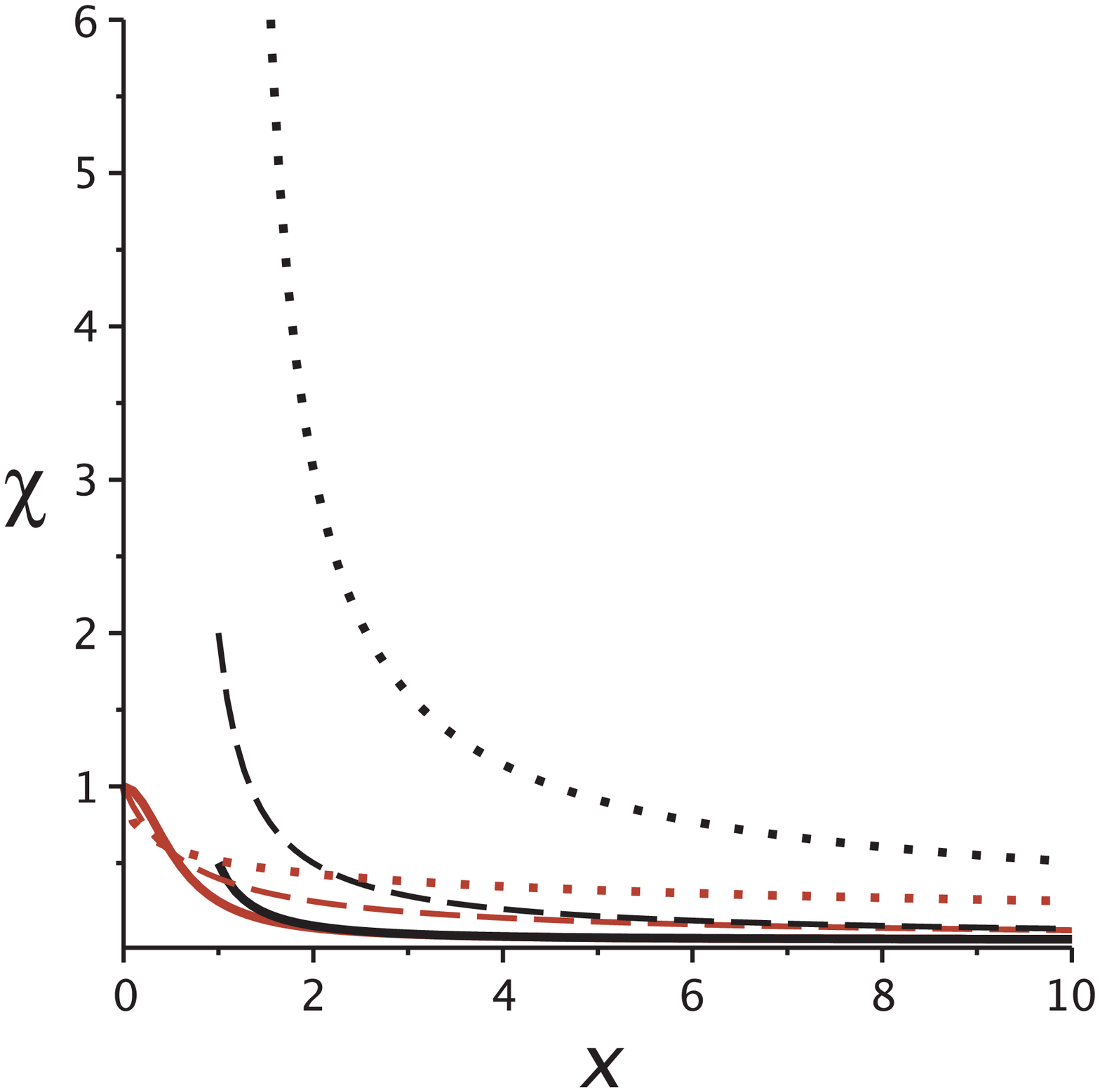} \hfill%\hspace{1em}
\includegraphics[width=0.33\textwidth]{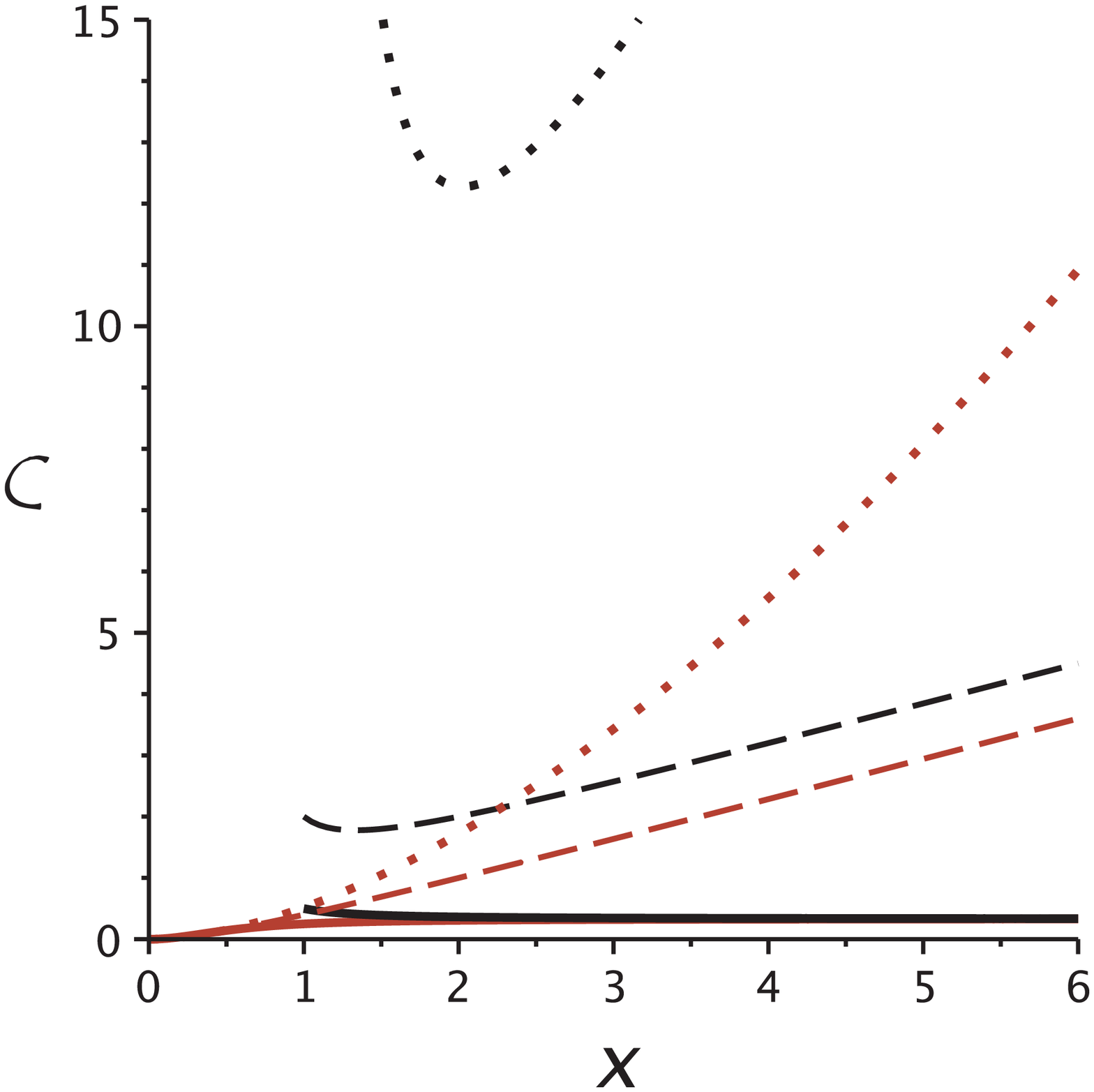} \hfill%\hspace{1em}
\includegraphics[width=0.33\textwidth]{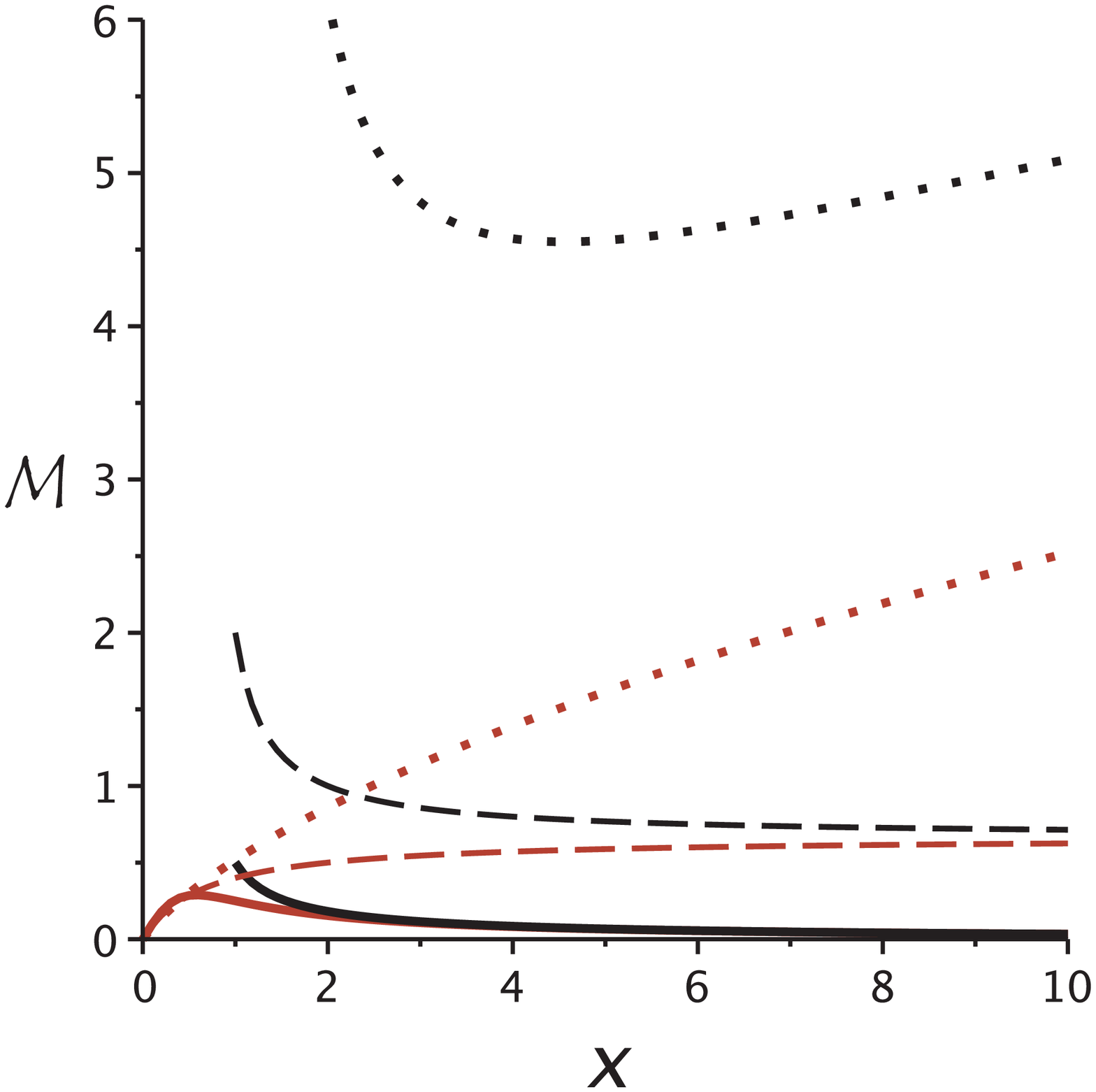}
\centerline{\hspace{5mm}{(a)} \hspace{15em} {(b)} \hspace{15em} {(c)}}
\caption{\label{Fig4} (Color online) Scaling functions for the isothermal
susceptibility (a), heat capacity (b), and
magnetocaloric coefficient (c). Light curves, brown online,
$T>T_\mathrm{c}$: $\chi_+(x)$, ${\cal C_+}(x)$, ${\cal M_+}(x)$. Black
curves, $T<T_\mathrm{c}$: $\chi_-(x)$, ${\cal C_-}(x)$, ${\cal M_-}(x)$.
Solid, dashed and dotted curves correspond to  the values of $q$ and
$\lambda$ of the free energy scaling functions of figure~\ref{Fig2}.}
\end{center}
\end{figure}

In figures~\ref{Fig4} we show the behaviour of the scaling functions
for thermodynamic observables that characterize the response of a
system to an external action, i.e., the isothermal susceptibility [figure~\ref{Fig4}~(a)], heat capacity [figure~\ref{Fig4}~(b)], and
magnetocaloric coefficient [figure~\ref{Fig4}~(c)]. The values of
$q$ and $\lambda$, for which the curves are plotted are the same as
those for the free energy scaling functions in figure~\ref{Fig2}: they
reflect the limiting behaviour at some marginal value
$\lambda_\mathrm{c}(q)$. At $1\leqslant q \leqslant 2$ and $\lambda > \lambda_\mathrm{c}(q)$, the
phase transition remains the second order but the critical exponents
do not depend on $\lambda$ any more, the scaling function does not
depend on $\lambda$ either. However, for $q>2$,
$\lambda>\lambda_\mathrm{c}(q)$, the phase transition turns to the first order
and the scaling regime does not hold any more. In turn, in the
region below $\lambda_\mathrm{c}$, the exponents acquire $\lambda$-dependency,
so do the scaling functions, as is plotted in the figures.

\section{Conclusions} \label{V}

In this paper we were interested in an analysis  of the critical
behaviour of the Potts model on uncorrelated scale-free network.
In our previous work the list of critical exponents for the Potts model
in the second order phase transition regime was obtained \cite{Krasnytska13}. Here,
we complete quantitative characteristics of the universal features
by calculation of the amplitude ratios and scaling functions.
Our results are exact for the annealed scale-free network and
correspond to the mean field treatment of the quenched case.

We obtain general scaling functions for the order parameter,
entropy, the constant-field heat capacity, magnetic susceptibility
and the isothermal magnetocaloric coefficient near the critical
point. The comprehensive list of scaling functions and critical
amplitude ratios was obtained in different regions of $q$ and
$\lambda$. It was shown that the critical amplitude ratios  are
$\lambda$-dependent similar to the critical exponents, so $\lambda$
plays the role of a global parameter of the system.

%\clearpage

\section*{Acknowledgements}

This work was supported in part by FP7 EU IRSES projects No.
$269139$ `Dynamics and Cooperative Phenomena in Complex Physical and
Biological Media', No.~295302 `Statistical Physics in Diverse
Realizations' and by the Coll\`{e}ge Doctoral $02-07$
Statistical Physics of complex systems. It is my big pleasure to
thank Bertrand Berche and Yurij Holovatch  for useful comments and
discussions.

% \vspace{0.5cm}

\clearpage

\ukrainianpart

\title{Скейлінгові функції  та співвідношення амплітуд для моделі Поттса на нескорельованій безмасштабній мережі}
\author{М. Красницька\refaddr{label1,label2}}
\addresses{
\addr{label1} Інститут фізики конденсованих систем НАН України,
вул. І.~Свєнціцького, 1, 79011 Львів, Україна
\addr{label2} Інститут Жаня Лямура, CNRS/UMR 7198, Група
 статистичної фізики, Університет  Лотарингії, \\
 BP 70239, F-54506 Вандувр лє-Нансі,
 Франція}

\makeukrtitle

\begin{abstract}
\tolerance=3000%
Ми вивчаємо критичну поведінку $q$-станової  моделі Поттса на
нескорельованій безмасштабній мережі зі степеновою функцією
розподілу за ступенем вузлів  з показником загасання $\lambda$.
Попередні результати показали, що фазова діаграма моделі в площині
$q$, $\lambda$ в режимі фазового переходу другого роду містить три
області, кожна з яких характеризується різним набором критичних
показників. У даній роботі ми доповнимо ці результати знайшовши
аналітичні представлення скейлінгових функцій та співвідношення
критичних амплітуд у згаданих вище областях. Як і для знайдених
раніше критичних показників, виявляється, що скейлінгові функції та
співвідношення амплітуд є $\lambda$-залежними.  Таким чином, ми
подаємо повний опис критичної поведінки у новому класі
універсальності.
\keywords  модель Поттса, складні мережі, скейлінг, універсальність
%
%\pacs 64.60.ah, 64.60.aq, 64.60.Bd
\end{abstract}

\end{document}